%% file: loizides-qm2008.tex
\documentclass[12pt]{iopart}
\usepackage{epsfig}
\usepackage{cite}
\usepackage{hypernat}
\usepackage{color}
\usepackage{ifpdf}

\input{header.tex}
\pqmtrue

\begin{document}
%--------------------------------------------------------------------------
\title[Photon-tagged jet measurements in Pb+Pb with the CMS detector]
{Photon-tagged jet measurements in Pb+Pb collisions at $\snnbf$=5.5~TeV with the CMS detector}
\author{Constantin Loizides for the CMS collaboration}
\address{Massachusetts Institute of Technology, 77 Mass Ave, Cambridge, MA 02139, USA}
\ead{loizides@mit.edu}

\input{abstract.tex}

%\pacs{25.75.-q}

%Uncomment for Submitted to journal title message
%\submitto{\jpg}

%Comment out if separate title page not required
%\maketitle

%\normalsize

\input{document.tex}

%--------------------------------------------------------------------------
\end{document}

%% file: header.tex
\usepackage{epsfig}
\usepackage{cite}
\usepackage{hypernat}
\usepackage{color}
\usepackage{ifpdf}

\newif\ifcomment
\newif\ifack
\newif\ifpqm
\newif\ifprint
\printtrue
\newif\ifcmscr

\ifprint
 \usepackage[linkbordercolor={0 0 .8},%
             citebordercolor={.8 0 0},%
             urlbordercolor={.8 0 .8},%
             raiselinks=true,%
             pdfborder={0 0 1 [3]}]{hyperref}
\else
 \usepackage[pdftex,backref,pagebackref,colorlinks,%
             linkcolor=blue,citecolor=blue,anchorcolor=blue,%
             pagecolor=blue,urlcolor=red,a4paper]{hyperref}
\fi

\graphicspath{{./pics/}}

\def\ECAL{ECAL}
\def\HCAL{HCAL}
\def\wrt{with respect to}

\newcommand {\htodo}[1]    {}

\newcommand {\abs}[1]      {\ensuremath{\left| #1 \right|}}

\newcommand {\snn}         {\ensuremath{\sqrt{s_{\scriptscriptstyle{{\rm NN}}}}}}
\newcommand {\snnbf}       {\ensuremath{\mathbf{\sqrt{s_{\scriptscriptstyle{{\mathbf NN}}}}}}}

\renewcommand {\pt}        {\ensuremath{p_{\mathrm{T}}}}

\newcommand {\et}          {\ensuremath{E_{\mathrm{T}}}}
\newcommand {\etg}         {\ensuremath{E^{\gamma}_{\mathrm{T}}}}
\newcommand {\etj}         {\ensuremath{E^{\rm jet}_{\mathrm{T}}}}

\newcommand {\AuAu}        {Au+Au}
\newcommand {\CuCu}        {Cu+Cu}
\newcommand {\PbPb}        {Pb+Pb}
\newcommand {\pp}          {p+p}

\newcommand {\dAu}         {d+A}

\newcommand {\dphiaj}      {\ensuremath{\Delta \phi(\gamma,{\rm jet})}}

\newcommand {\gsim}        {\,{\buildrel > \over {_\sim}}\,}
\newcommand {\Ref}[1]      {Ref.~\cite{#1}}

\newcommand {\Fig}[1]      {Fig.~\ref{#1}}

\newcommand {\arxiv}[1]    {\href{http://www.arxiv.org/abs/#1}{\mbox{arXiv:#1}}}
\newcommand {\hrefurl}[1]  {\href{#1}{\url{#1}}}
\newcommand {\dd}          {\rm d}

\newcommand {\beq}         {\begin{equation}}
\newcommand {\eeq}         {\end{equation}}
\newcommand {\beqa}        {\begin{eqnarray}}
\newcommand {\eeqa}        {\end{eqnarray}}
\newcommand {\beqnn}       {\begin{equation*}}
\newcommand {\eeqnn}       {\end{equation*}}
\newcommand {\beqann}      {\begin{eqnarray*}}
\newcommand {\eeqann}      {\end{eqnarray*}}

%% file: abstract.tex
% $Id: abstract.tex 2296 2008-04-04 20:40:48Z loizides $

\begin{abstract}
Presented are the results of a detailed study for a complete simulation of the CMS detectors 
at the LHC in view of the expected modification of jet fragmentation functions in central 
\PbPb\ collisions at $\snn=5.5$~TeV compared to the vacuum~(\pp) case.
The study is based on $\gamma$-jet events, using the correlation between isolated 
high-transverse energy ($E_T>70$~GeV) photons and fully reconstructed jets, based 
on the information provided by the CMS calorimeters and silicon tracker.
\end{abstract}

%% file: document.tex
% $Id: document.tex 2296 2008-04-04 20:40:48Z loizides $

\begin{figure}[t!f]
\begin{center}
\resizebox{0.46\textwidth}{!}{\includegraphics{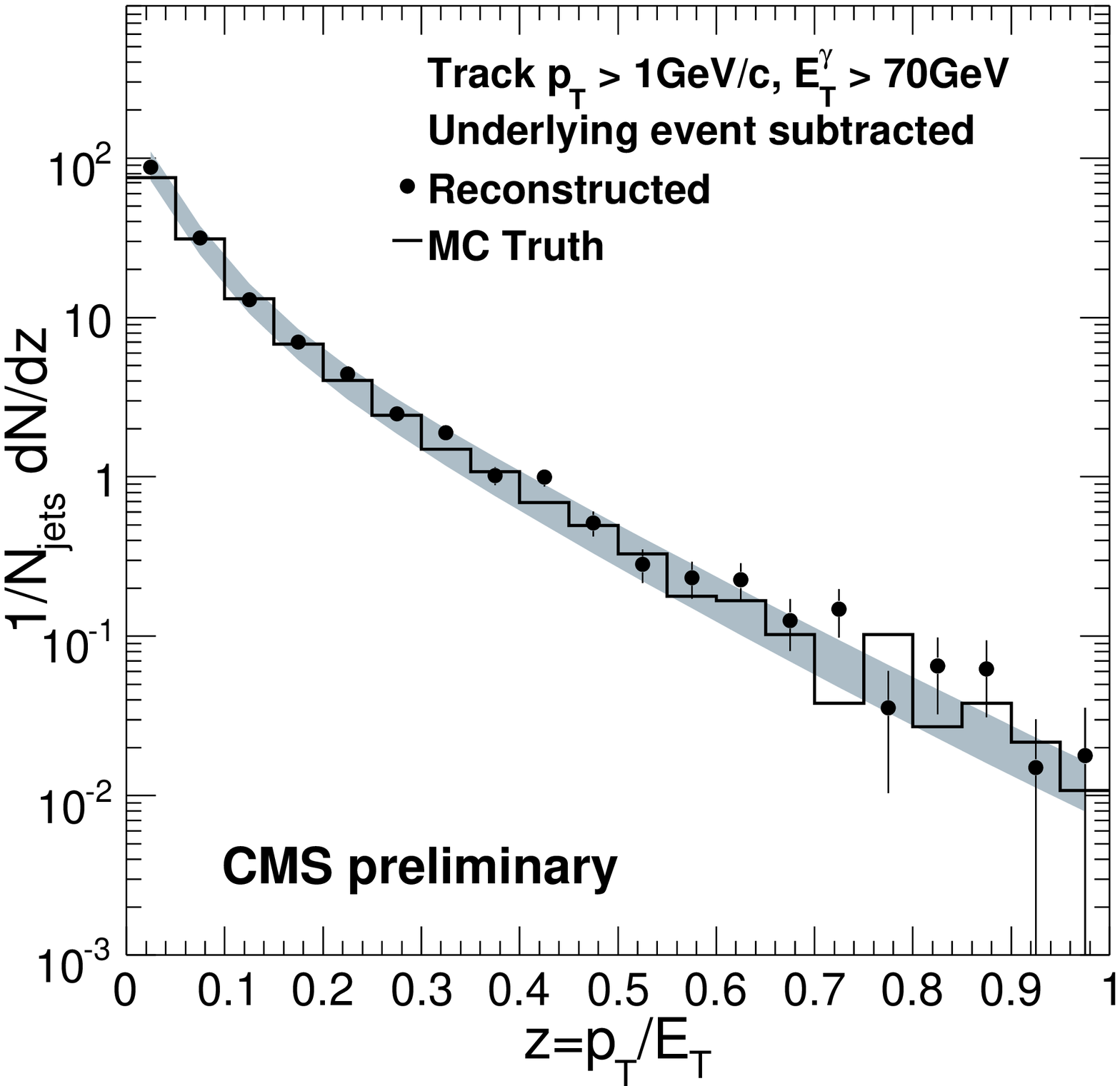}}
\hspace{0.05\textwidth}
\resizebox{0.46\textwidth}{!}{\includegraphics{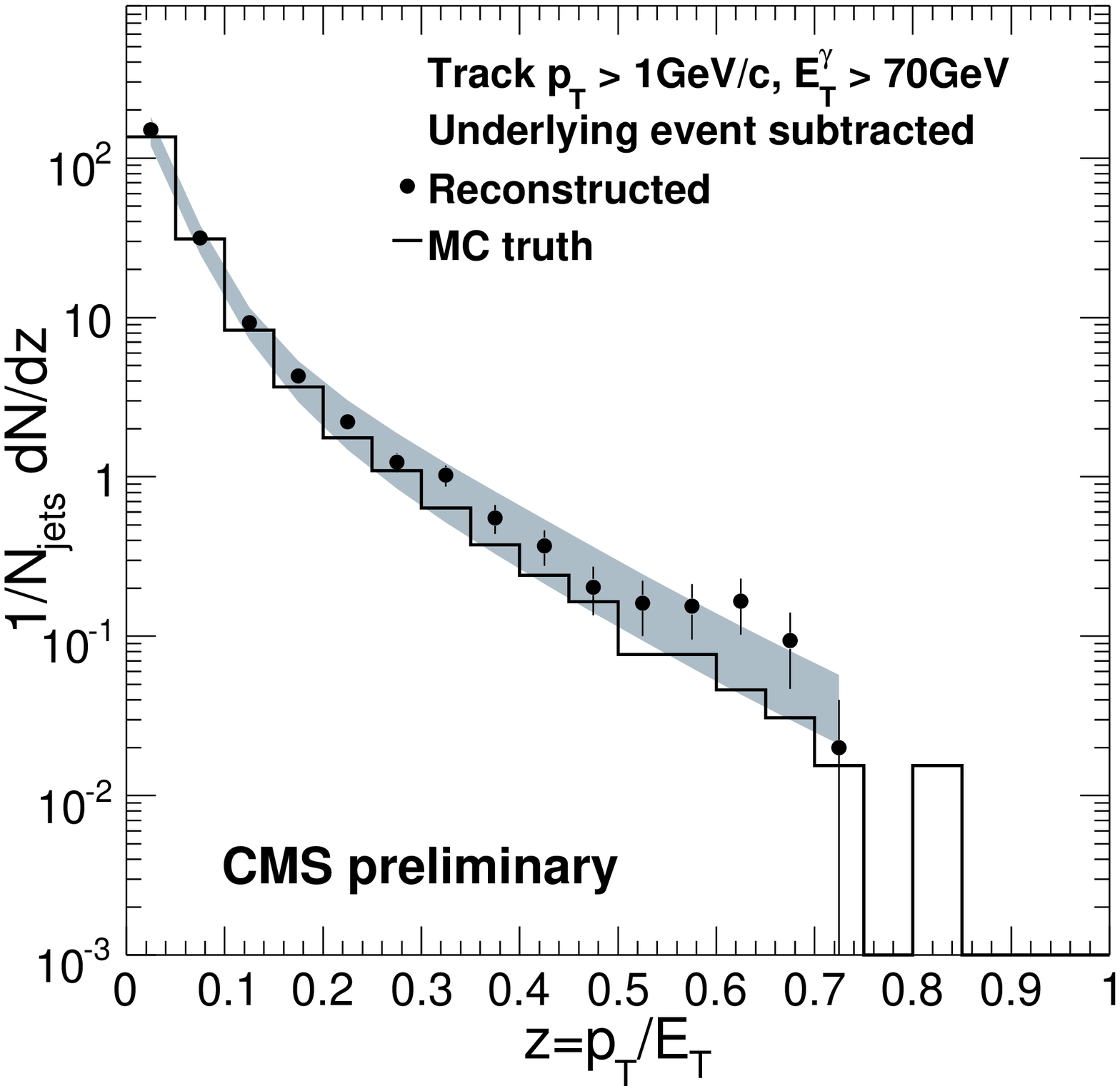}}
\caption{\label{fig:zpbpbsubffmc}
Reconstructed fragmentation function in \PbPb\ collisions~(symbols) 
compared to MC truth~(line) for the unquenched~(left) and quenched~(right) scenario.
The estimated systematic error of the measurement is represented as the shaded band.}
\end{center}
\end{figure}

%--------------------------------------------------------------------------
\ifcmscr
\section{\label{sec:intro}Introduction}
\fi
%--------------------------------------------------------------------------
One of the key results from the study of high-energy nuclear~(\AuAu\ and \CuCu, $\snn=200$~GeV) 
collisions at the Relativistic Heavy-Ion Collider~(RHIC) is the observation of the strong 
suppression of leading hadron yields at transverse momenta $\pt \gsim 5$~GeV/$c$, 
compared to expectations based on \pp\ and \dAu\ collisions at the same collision 
energies~\cite{whitepapers}. This observation is called ``jet-quenching'' and commonly explained 
by assuming that the hard partons lose a large fraction of 
%produced at mid-rapidity in high-energy nucleus--nucleus collisions 
their initial energy due to interactions with the surrounding strongly-interacting medium.
\ifpqm
The extreme magnitude of the suppression~(up to factor 5) %in single-inclusive hadron spectra) 
makes the interpretation of the inclusive measurement difficult, as the observed remaining yield 
of high-$\pt$ hadrons is likely to be dominated by emission from the surface of the collision 
region~\cite{Dainese:2004te,Eskola:2004cr}, and therefore allows a variety of qualitative different 
models to describe the data quantitatively equally well~\cite{Adare:2008cg}.
\fi
The collisional and radiative energy loss of the hard partons should lead 
to characteristic changes in the jet fragmentation pattern~\cite{Salgado:2002cd}, but a direct 
measurement of jets in central nucleus--nucleus collisions is not feasible at RHIC energies. 
\PbPb\ collisions at $\snn=5.5$~TeV at the Large Hadron Collider~(LHC) will allow one 
to study the jet quenching phenomenon in more detail. Due to the very large cross-sections for 
hard parton-parton scattering, a significant fraction of the produced jets will have enough 
transverse energy to individually stick out of the underlying heavy-ion background. Therefore
identified jets can be used to characterize the initial partons event-by-event.
The CMS detector will collect large event samples for the study of such rare probes, 
thanks to the large geometrical coverage of the electromagnetic~(\ECAL) and hadronic~(\HCAL) 
calorimeters~($\abs{\eta}<3$) and of the charged-particle tracking~($\abs{\eta}<2.5$), 
both fully covering the azimuthal angle. 
Equally important is the capacity of its triggering and data acquisition systems to inspect 
every \PbPb\ collision, giving priority to the storage of rare events~\cite{hitdr}. The expected 
statistical significance of the event samples, 
%good enough to study jets of transverse energies far beyond $100$~GeV, 
will allow us to study in detail even the $\gamma$-jet channel. This channel 
is particularly interesting since the direct photon is not affected by the presence of the medium, 
and thus can be used as an unambiguous tag of the away-side parton, contrary to the case of 
back-to-back di-jets~\cite{Wang:1996yh}. %,Arleo:2006xb}
The main idea of the analysis is to use the transverse energy of the direct photon~($\etg$) as 
an estimate of the $\etj$ of the away-side jet, and thus to avoid the precise measurement of 
the absolute $\etj$. Hence, the jet fragmentation function ${1}/{N_{\rm jets}}\,\dd N/\dd z$, 
the normalized distribution of fractional particle transverse momenta relative to $\etj$, 
can be approximated by using $z = \pt/\etg$ for particles associated with the 
jet~(\Fig{fig:zpbpbsubffmc}). 
Of course, to be useful as a tag, the direct photon should emerge back-to-back \wrt\ the 
away-side jet. This topology is disturbed by initial and final state radiation. Therefore, a 
cut on the opening angle between the photon and the emerging jet is applied. 
%, requiring the angle to be greater than $172^\circ$ in azimuth. 

In this report, we present the jet fragmentation functions for photon-tagged jet events
with $\etg>70$~GeV, reconstructed from the simulated response of the CMS detector 
systems for central \PbPb\ collisions. 
\ifcmscr
The studies are performed for two extreme scenarios.
The ``unquenched'' scenario, using the
PYTHIA~\cite{pythia} generator, does not include parton energy loss. Consequently, it leads 
to larger hadron yields at high $\pt$ than commonly expected and therefore challenges the
photon selection performance due to larger decay backgrounds. 
The ``quenched'' scenario is simulated with the PYQUEN~\cite{pyquen} generator, which has been 
tuned based on the RHIC results. PYQUEN presents an extreme case, as the energy lost by the partons 
is largely radiated out of the jet cone around the parton axis. The lower total energy remaining 
close to the original parton direction challenges the performance of the jet finding algorithm. 
\else
The studies are performed for two extreme scenarios, where PYQUEN~\cite{pyquen} and 
PYTHIA~\cite{pythia} are used to generate the QCD signal and background channels with 
and without jet quenching, respectively. 
\fi
In both scenarios, 
HYDJET~\cite{hydjet} is used to model the underlying heavy-ion event. For this study, the $10$\% 
most central \PbPb\ collisions are selected by the impact parameter of the lead nuclei, yielding an 
average mid-pseudorapidity density of about 2400~(2200) charged particles in the quenched~(unquenched) case.
In total, 4000 $\gamma$-jet events in the CMS acceptance for $\etg>70$~GeV and \mbox{$\abs{\eta^\gamma}<2$} 
with $\dphiaj>172^\circ$ and about 40000~(125000) QCD background events for the 
quenched~(unquenched) case are simulated. This corresponds to the expected yields for one 
running year of \PbPb\ data taking with an integrated luminosity of 0.5~nb$^{-1}$.
More details can be found in~\Ref{moreinfo}.

\begin{figure}[t!f]
\vspace{-0.25cm}
\begin{minipage}[t]{0.48\textwidth}
\begin{center}
\includegraphics[width=0.95\textwidth]{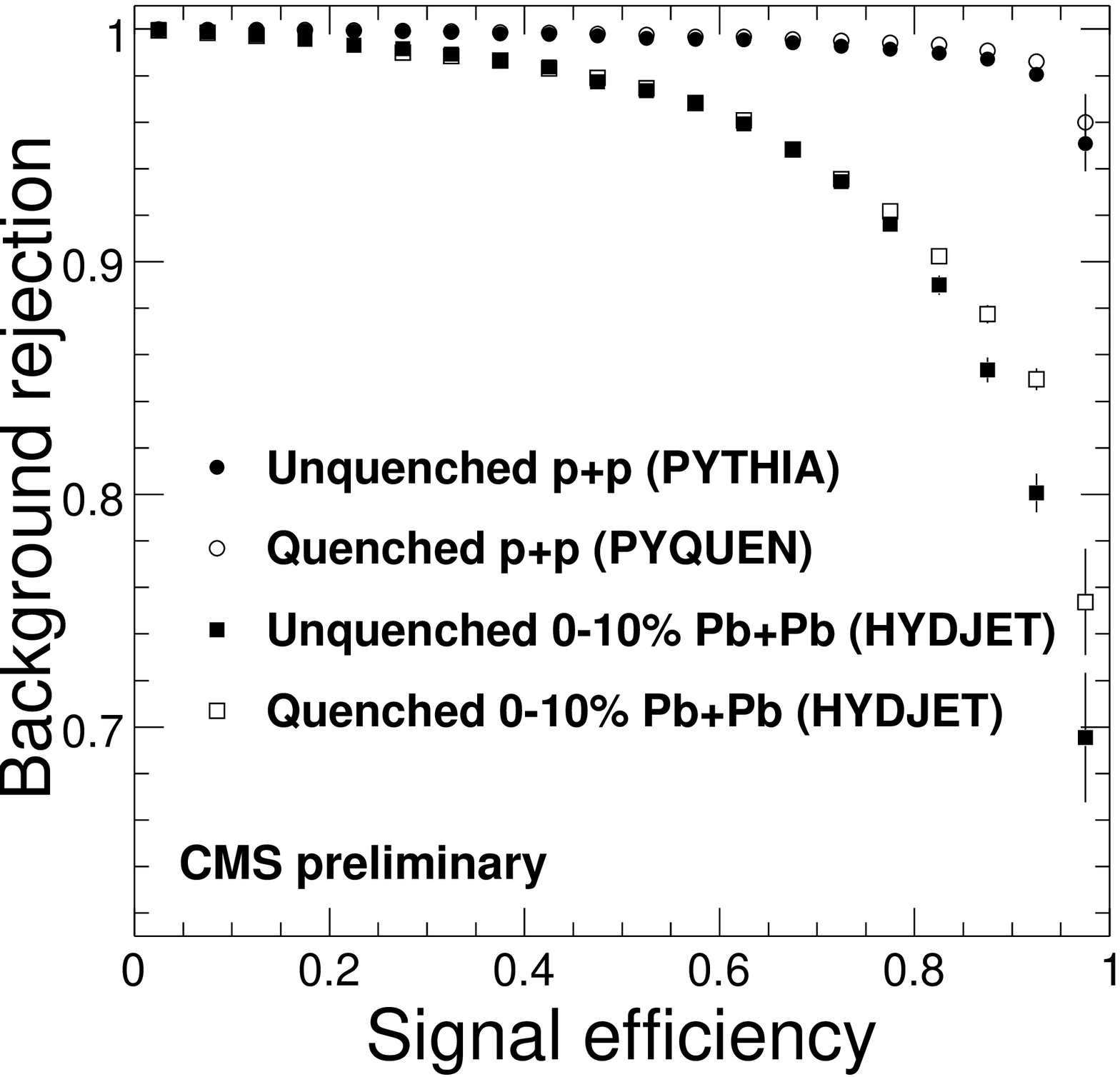}
\caption{\label{fig:effvsback}
Background rejection versus signal efficiency for the identification of isolated photons 
in different systems. The coefficients and cuts are obtained for the corresponding unquenched 
sample and applied to the quenched one.}
\end{center}
\end{minipage}
\hfill
\begin{minipage}[t]{0.48\textwidth}
\begin{center}
\includegraphics[width=0.95\textwidth]{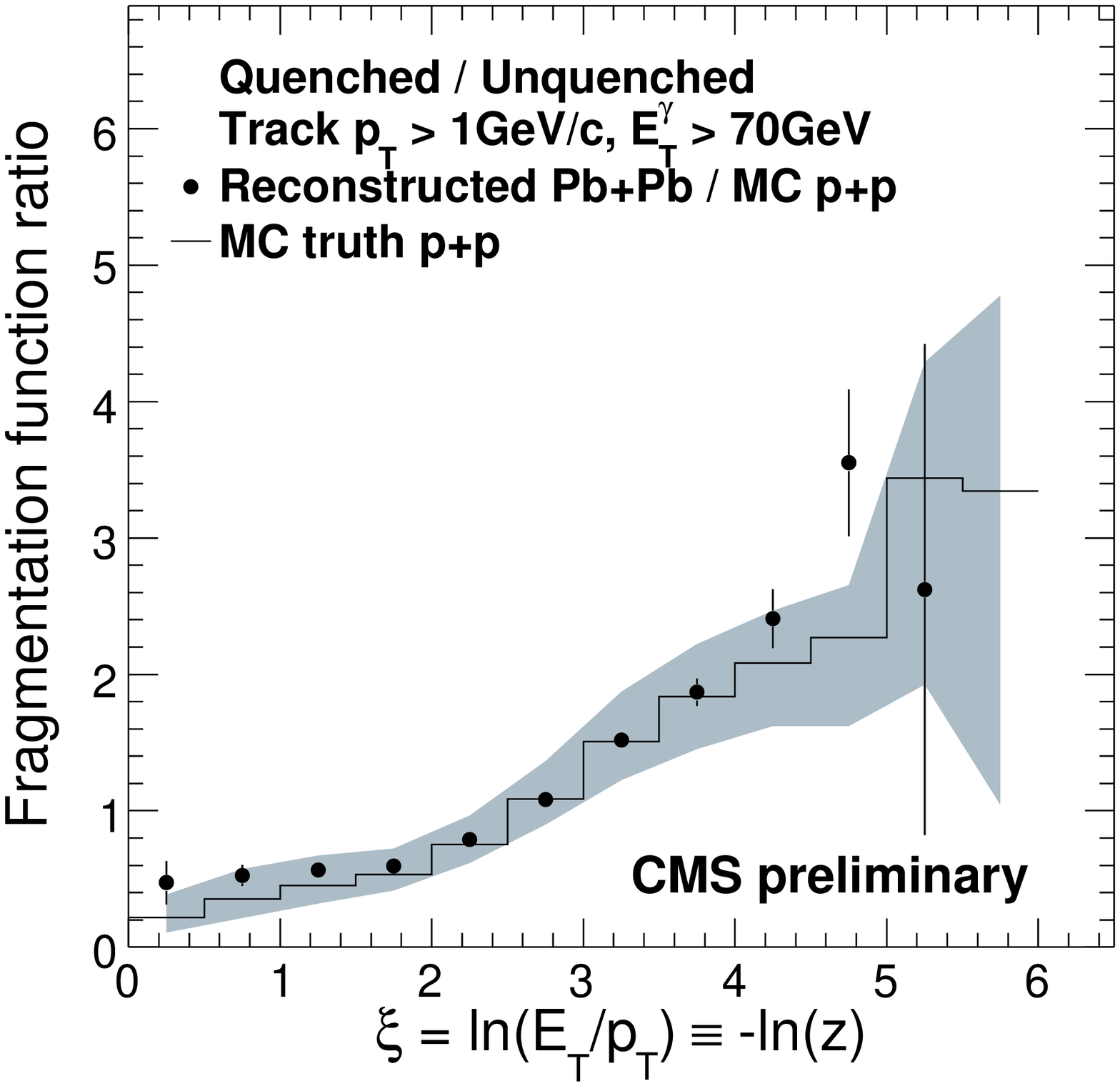}
\caption{\label{fig:quenunquen}
Ratio of reconstructed~(symbols) and MC truth~(line) quenched fragmentation 
function over unquenched MC truth. The estimated systematic error of the 
measurement is represented as the shaded band.}
\end{center}
\end{minipage}
\end{figure}

%--------------------------------------------------------------------------
\ifcmscr
\section{Event reconstruction}
\fi 
%--------------------------------------------------------------------------
Jet reconstruction using the \ECAL\ and \HCAL\ calorimeters is performed by an iterative cone 
algorithm with a cone size of $R=0.5$ in the $\eta$-$\phi$ plane modified to subtract 
the underlying soft background on an event-by-event basis~\cite{hitdr}. 
%The algorithm subtracts from each jet the background energy due to the underlying
%\PbPb\ event and electronic noise. The performance of this algorithm in documented 
%in \Refs{hitdr,Kodolova2007}.
Tracks are reconstructed using seeding from one hit on each of the three layers
of the silicon pixel detector~(resulting in a geometrical acceptance of $80$\%), 
with an extension of the standard tracking algorithm used for \pp~\cite{hitdr}.
In this high-multiplicity environment, %~($\av{\dncde=2400}$), 
an algorithmic tracking efficiency of about $70$\% is achieved near midrapidity, 
for $\pt>1$~GeV/$c$, with a few percent fake track rate. %Tracks are reconstructed 
%with excellent momentum resolution, $\Delta \pt/\pt<1.5$\% for $\pt<100$~GeV/$c$. 
%The resolution of the transverse track impact parameter at the event vertex is better than 
%$50$~$\mu$m at $\gsim1$~GeV/$c$, and improves to $20$~$\mu$m at high $\pt$~(above $10$~GeV/$c$).  
High-$\et$ isolated photon reconstruction, newly developed for this analysis, proceeds in three 
steps. At first, photon candidates are obtained by reconstructing clusters of hits measured
in the \ECAL\ using the Island clustering algorithm~\cite{cmsptdrv1}.
Second, for each photon candidate, the information provided by several shape variables in the 
\ECAL\ associated with the candidate is examined. Third, together with the information 
from the \HCAL\ and the tracker, we determine if a given photon candidate is an isolated photon.
The combined information of these variables forms a three-dimensional space, in which optimal
rectangular cuts are obtained~(\Fig{fig:effvsback}). %using the TMVA~\cite{tmva} package.
%The resulting background rejection power as a function of signal efficiency is shown in 
%\Fig{fig:effvsback} for quenched and unquenched \pp\ and $0-10$\% central \PbPb\ events. 
%It should be emphasized that for each collision system the coefficients and cuts of the 
%corresponding unquenched training samples are used to study the quenched case. 
The working point for this analysis is set to 60\% signal efficiency, leading to a background
rejection of about 96.5\%, and to a signal-to-background ratio of $4.5$ for $0-10$\%~central quenched 
\PbPb. On average, for $\etg>70$~GeV, the transverse energy resolution for isolated photons is about 
$4.5$\%, and the spatial resolution in $\eta$ and $\phi$ is better than 0.005.


%--------------------------------------------------------------------------
\ifcmscr
\section{Results}
\fi
%--------------------------------------------------------------------------
To construct the fragmentation function, reconstructed isolated photons 
are selected and correlated with reconstructed back-to-back calorimeter jets. 
An isolated photon with $\etj>70$~GeV and $\abs{\eta^\gamma}<2$ is associated 
with the highest $\etj$ calorimeter jet in the event provided $\dphiaj > 172^\circ$
is fulfilled. 
Since we obtain the charged particle information from the tracker, the away-side 
jet~($R=0.5$) associated with the photon needs to be contained in the tracker 
acceptance of $\abs{\eta}<2.5$. Therefore, we require the reconstructed jet axis 
to be within $\abs{\eta}<2.0$ to avoid edge effects at the limit of the tracker 
acceptance. A minimum $\etj$ cutoff of $30$~GeV is applied in order to ensure 
that the reconstructed calorimeter jet corresponds to the away-side parton.
For the selected photon-jet pairs, reconstructed charged particles that lie 
within the $R=0.5$ cone size around the reconstructed jet axis are selected. 
The raw fragmentation function is constructed correlating the transverse energy of 
the photon, as a measure of the parton transverse energy, with the reconstructed 
transverse momentum of the tracks in the cone.
%In central \PbPb\ events, special care needs to be taken to estimate the underlying event
%contribution from the copious soft background of the \PbPb\ events. 
The underlying event contribution is estimated by using the momentum distributions of tracks 
within a $R=0.5$ radius perpendicular in the $\phi$ to the reconstructed jet axis and subtracted from 
the raw distribution. %Neglecting higher order effects like flow, 
The reconstructed fragmentation functions are overlaid with the MC truth determined at 
generator level using the true parton \et\ and direction for the selection of 
particles~(\Fig{fig:zpbpbsubffmc}). 
Essentially four main sources are found to contribute to the systematic differences
between the reconstructed and true fragmentation functions and added in quadrature
to obtain the total systematic error:
\begin{enumerate}
\item QCD jet fragmentation products misidentified as isolated photons~($\sim15$\%).
\item Association of a wrong/fake jet on the away-side of the isolated photon~($\sim10$\%).
\item Uncertainties in the charged particle reconstruction efficiency correction~($\sim10$\%). 
\item Biases due to lower jet reconstruction efficiency at lower jet $\et$~($\sim30$\%).
\end{enumerate}
%The true fragmentation functions extracted from the generator level information 
%are very well reproduced by the reconstructed fragmentation functions based on the 
%photon, jet and charged particle reconstruction, for both scenarios.
The overall capability to measure the medium-induced modification of jet fragmentation functions 
in the $\gamma$ jet channel can be illustrated by comparing the fully reconstructed quenched 
fragmentation function to the unquenched MC truth distribution~(\Fig{fig:quenunquen}). The change 
in the fragmentation function between the unquenched and quenched case provides the scale against 
which the estimated uncertainties should be compared.
%where $\xi=-\ln z=\ln (\et/pt)$ as used in \Ref{Borghini:2005em}.

In summary, we have shown that $\gamma$-jet events can be used to study quantitatively the 
dependence of high-$\pt$ fragmentation on the medium. For a data set corresponding to
one nominal year of CMS \PbPb\ running, the expected statistical and systematic uncertainties 
should be small enough that the measurements will be sensitive to the foreseeable changes in the 
fragmentation functions relative to parton fragmentation in vacuum. 
%Therefore this measurement will allow a quantitative test of proposed mechanisms for parton energy 
%loss in the medium, testing fundamental properties of the high-density QCD medium produced in 
%high-energy nuclear collisions.

\ifack
%--------------------------------------------------------------------------
%\section*{Acknowledgments}
%--------------------------------------------------------------------------
\medskip
\fi

%--------------------------------------------------------------------------
\ifcmscr
\else
\section*{References}
\fi
%--------------------------------------------------------------------------